\documentclass[preprint,preprintnumbers,amsmath,amssymb]{revtex4}

\usepackage{amsmath}
\usepackage{amssymb}
\usepackage{graphicx}
\usepackage{dcolumn}
\usepackage{bm}

\def\be{\begin{eqnarray}}
\def\ee{\end{eqnarray}}
\def\ba{\begin{array}}
\def\ea{\end{array}}

\begin{document}

\title{Screening Model of Magnetotransport Hysteresis Observed in
 Bilayer Quantum Hall Systems}

\author{Afif Siddiki}, %
\author{Stefan Kraus} %
\author{Rolf R. Gerhardts}
\address{Max-Planck-Institut f\"ur Festk\"orperforschung,
Heisenbergstrasse 1, D-70569 Stuttgart, Germany}%


\begin{abstract}
We report on theoretical and experimental investigations
of a novel hysteresis effect that has been observed on the
magnetoresistance of quantum-Hall bilayer
systems. Extending to these system a recent approach, based on the
Thomas-Fermi-Poisson nonlinear screening theory and a local
conductivity model,  we are able to explain
the hysteresis as being due to screening effects such as the
formation of ``incompressible strips'', which hinder the electron density
in a layer within the quantum Hall regime to reach its equilibrium
distribution.

\end{abstract}

\maketitle

\section{\label{sec:1} Introduction}

Recent  scanning force microscope
experiments \cite{Ahlswede01:562,Ahlswede02:165} and subsequent
theoretical work \cite{Guven03:115327,siddiki2004} have shown that
screening effects, notably the occurrence of ``incompressible
strips'' \cite{Chklovskii92:4026,Siddiki03:125315}, are very important
for the understanding of the Hall-field and current distribution as
well as the high precision of the low-temperature integer quantized
Hall (QH) effect
in narrow Hall samples. Since Coulomb interactions should become
important within and between
the layers of an electron bilayer system showing the  {\sl drag
effect} \cite{Gramila91:1216,Zheng93:8203,Bonsager97:10314} at very low
temperatures, we extended the
theory  of Ref.~\cite{siddiki2004} to the bilayer case and
calculated the Hall and longitudinal resistances for density-matched
and mismatched systems. Magnetoresistance (MR) measurements on separately
contacted bilayers,
which were performed to check the results, showed pronounced
hysteresis effects, similar to results reported previously for
hole \cite{tutuc}  and electron \cite{pan} double layers. In contrast to
this previous work, our approach makes the origin of the hysteresis
rather evident: the peculiar nonlinear screening properties leading to the
occurrence of ``incompressible strips'' in the
QH regimes of the individual layers.

\section{The Experiments} \label{experiments}
We measured the MRs of both layers as functions of the applied
perpendicular magnetic field and  the sweep direction  for matched
and mismatched densities at a fixed base temperature ($T=270$mK)
within the linear response regime ($I\sim 50$nA).
\begin{figure}[t] \centering
\includegraphics[width=0.6\linewidth,angle=0]{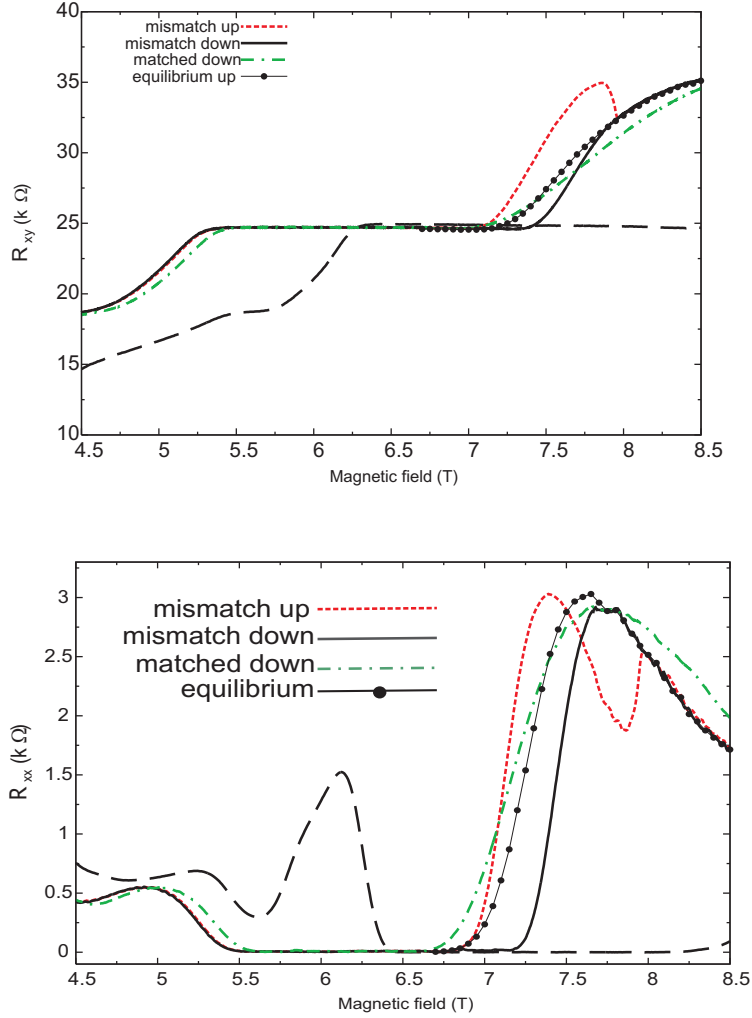}
\caption{\label{fig:experiment1} Measured Hall (upper panel) and
longitudinal (lower panel) resistances for the top layer
and the bottom layer (long-dashed lines, only for the mismatched case and
one sweep direction). No hysteresis
is observed for
the matched case (green dashed-dotted line) and the
``equilibrium
measurement'' (solid line with circles), with $n^{T}/n^{B}=0.84$.}
\end{figure}
\begin{figure}[h] {\centering
  \includegraphics[width=0.9\linewidth,angle=0]{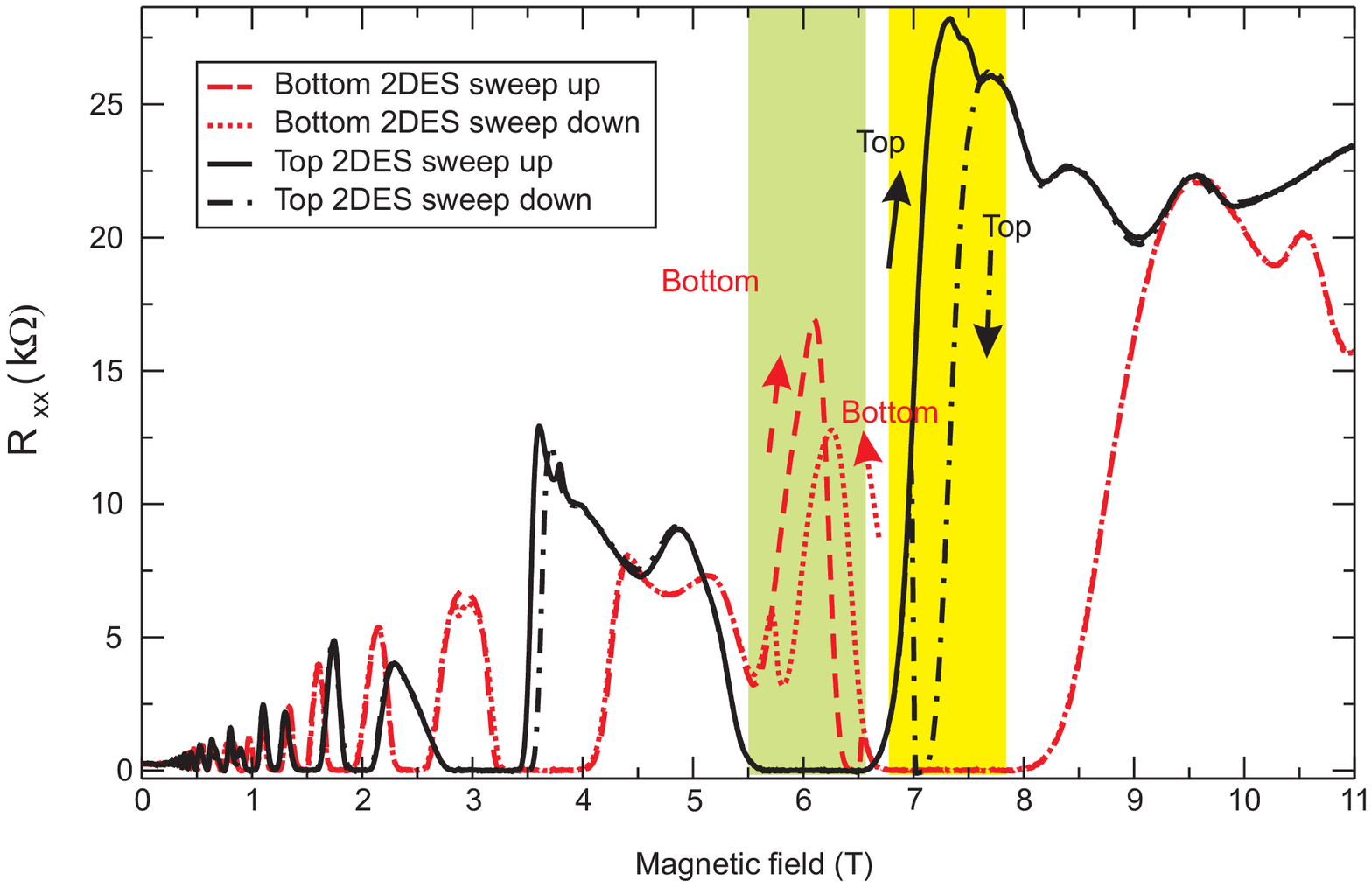}
\caption{ \label{fig:exphysteresis}Longitudinal resistances  versus
magnetic field,  for both layers and two sweep directions with
$n^{T}_{el}/n^{B}_{el}=0.82$.
 }}
\end{figure}
We also performed ``equilibrium'' measurements, where the system was
heated up to $\sim\!\!10\,$K and cooled down again at each magnetic
sweep step, in the magnetic field interval where hysteresis is
observed. The samples are $15\,$nm wide GaAs/AlGaAs double quantum
well structures grown by MBE and  capped by top and bottom
gates that control the electron densities of the layers. Two
2DES are confined by $\delta-$ doping remote Si donors and are
separated by a spacer of thickness $h=12\,$nm. For such a
separation the bilayer system is electronically decoupled, i.e.  electron
tunneling between the layers ($R_{tunne1}>100 $M$\Omega$) is not possible
and the system  can be described
by two different electrochemical potentials.  Separate Ohmic
contacts to the two layers are realized by a selective depletion
technique \cite{Eisenstein90:2324}. The samples were processed into
$80\,\mu$m wide and $880\,\mu$m long Hall bars, with grown densities
in the range $1.5-2.5\times10^{15}\,$m$^{-2}$, the mobility is
$100\,$m$^{2}$/Vs per layer.

Figure~\ref{fig:experiment1}  shows  Hall  and
 longitudinal resistances of the top layer  versus magnetic field
 strength, measured under different conditions. As a reference, the
 resistances of the bottom layer (long-dashed lines) are also shown
 for one sweep  direction in the mismatched case.  In the
 ``equilibrium measurements'' (solid
 lines with symbols) explained above and also in the case of matched
densities the resistances are insensitive to the sweep
 direction. Note that the QH plateau for matched densities is narrower
 than that of the mismatched system in the ``equilibrium measurement''.
 For the density mismatched (and non-equilibrium) case, the
data were taken at a sweep rate $0.01\,$T/min and the base
temperature is always kept at $270\,$mK. Apparently the resistances
of the top layer follow different traces for the up- (red dotted lines)
and the down- (black solid lines) sweep.

Figure \ref{fig:exphysteresis} shows that, for density-mismatched
bilayers, hysteresis occurs in both layers under comparable
conditions. The hysteresis shown in Fig.~\ref{fig:experiment1} occurs
in the top layer near the high-$B$ edge of the $\nu=2$ QH plateau
(plateau region $5.5\,$T$\lesssim B \lesssim 6.7\,$T). It
occurs in a $B$-interval well within the  $\nu=2$ QH plateau
($6.8\,$T$\lesssim B \lesssim 7.9\,$T) of the bottom layer.
Similarly, the MR of the bottom layer shows hysteresis at the low-$B$
edge of its  $\nu=2$ QH plateau, in a $B$-interval well within the
 $\nu=2$ QH plateau of the top layer. No hysteresis is observed in
magnetic field intervals in which the other layer is in the normal
state, with finite longitudinal resistance.  A
less pronounced repetition of these features is observed
in the $\nu=4$ plateau regimes.

\section{\label{sec:2} The Model}

We model the electron bilayer system as a series of parallel
charged planes, perpendicular to the $z$-direction, translation
invariant in the $y$-direction, and confined to $|x|<d$. The
bottom and top 2DESs with number densities $n^{B}_{\rm el}(x)$ and
$n^{T}_{\rm el}(x)$ lie in the planes $z=z_B\equiv 0$ and $z=z_T
\equiv h$, respectively. Ionized donor layers with number
densities $n^{T}_{0}$ and $n^{B}_{0}$ are assumed at $z=-c$ and
$z=h+c$, and a top gate at $z=h+c+f$, allowing for a density
mismatch even with $n^{T}_{0}=n^{B}_{0}=n_{0}$,  is simulated by
ionized donors with number density $
n_{g}(x)=n_{g}^{0}\cosh(\frac{5}{8 \pi}(x/d)) \label{gate}$ at
$z=c+h+f\equiv z_{g}$. Solving Poisson's equation with the
boundary condition $V(x=\pm d,y,z)=0$, we obtain from the charge
densities $\mp e n_j(x)$ in the plane $z=z_j$ as contribution to
the potential energy of an electron at position $(x,y,z)$:
\be V^{j}(x,z)=\pm  \frac{2e^2}{\bar{\kappa}} \int_{-d}^{d}
dx'\, K(x,x',z,z_j)\, n_j(x'), \label{eq:V_Hartree} \ee
with the kernel \cite{Siddiki03:125315}
$ K(x,x',z,z_j)=- \ln([c^2+\gamma^2]/[s^2+\gamma^2])$, where
$c=\cos(\pi[x+x']/4d)$, $s=\sin (\pi[x-x']/4d)$, and
$\gamma=\sinh(\pi[z-z_j]/4d)$. We write the total potential energy of
an electron as
\be \label{vbplusvt}
V(x,z)=V^B(x,z)+V^T(x,z)\,, \ee
where $V^B(x,z)$ is the sum of the potentials created by  bottom
electron and donor layer, while $V^T(x,z)$ is the sum of the
potentials due to the top electron, donor, and gate layers. The
electron number densities in the layers are, within the
Thomas-Fermi approximation (TFA),
 \be n^j_{\rm el}(x)=\int
dE\,D(E)\,f([E+V(x,z_j)-\mu^{\star}_j]/k_{B}T) \label{tfaed}\ee
with $j=B$ or $T$ and $D(E)$ the (collision-broadened) Landau density
of states. This completes the self-consistency scheme
\cite{Guven03:115327,siddiki2004}. In the practical calculations
we first decoupled the layers, replacing $ V(x,z_j)$ by $ V^j(x,z_j)$,
and solved the single electron-layer problem for bottom and top system
separately. There it is convenient to fix the edges $x=\pm b$ of the
electron density profile in the limit of zero magnetic field and
temperature \cite{Guven03:115327,siddiki2004}, which fixes the average
density (and  $\mu^{\star}_{B,T}$). With the converged results at finite
$T$ and $B$, we treat the inter-layer coupling iteratively.

To describe the density-mismatched case, we
add more electrons to the top layer by setting
$V_{0}=n_{g}^{0}/n_{0}$ to a positive value and keeping the
depletion length $d-b$ fixed.  We
scale energies by the average Fermi energy
$E^{*}_{F}=(E^{T}_{F}+E^{B}_{F})/2$ , e.g.
$\hbar\omega_{c}/E^{*}_{F}=\Omega_{c}/E^{*}_{F}$.

\begin{figure}[t]
{\centering
\includegraphics[width=1.0\linewidth]{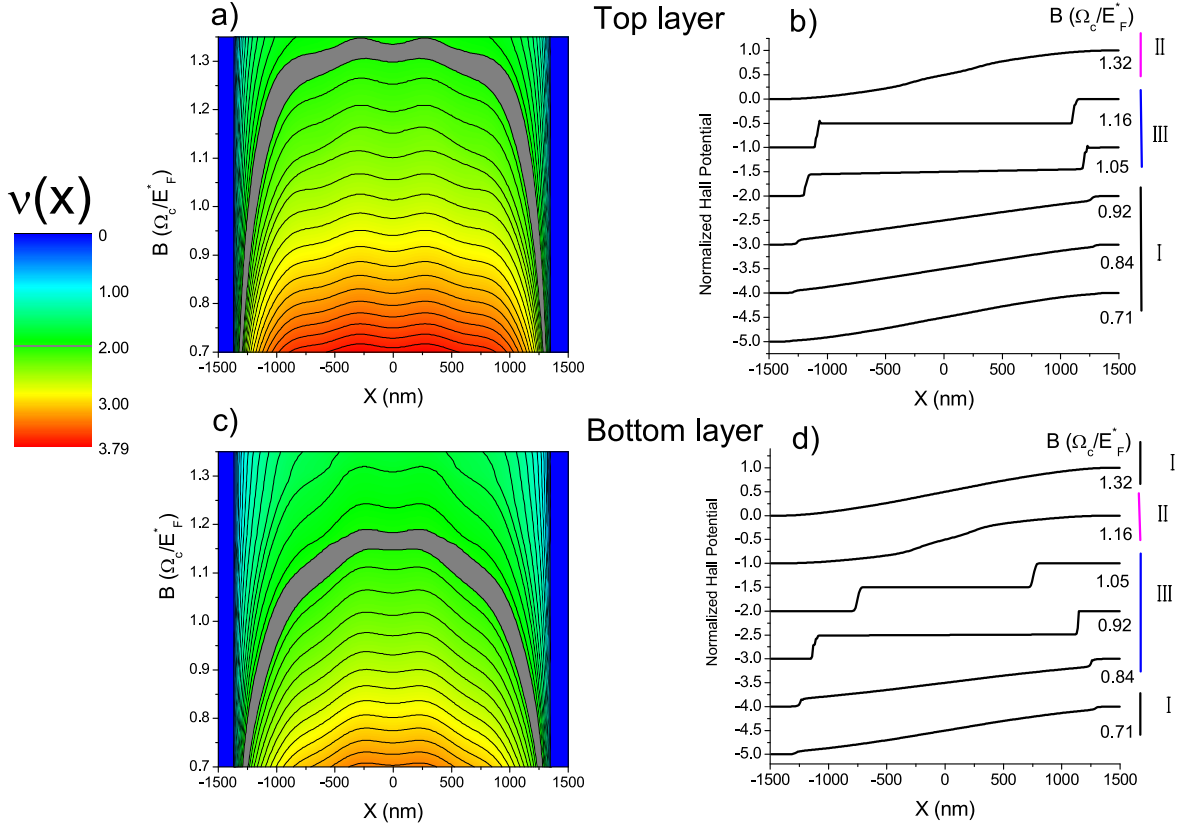}
\caption{Density profiles (color scale) across the sample as a
function of magnetic field (left panel) and (normalized) Hall
potential profile of both layers for selected $B$ values (right
panel). The density mismatch is obtained by setting
$V_{0}/E_{0}=0.05$ which results in $n^{B}_{el}/n^{T}_{el}=0.84$
at a low temperature $k_{B}T/E_{F}^{*}=0.0124$ for fixed
$b/d=0.9$. Here $E_{0}(=2\pi e^2 n_0 d/\bar{\kappa})$ is the
pinch-off energy which defines the minimum of the bare confinement
potential.
 \label{fig:nonequdens}}}
\end{figure}

 Next we implement in each electron layer a quasi-local
model  to impose a current, following the lines of
Ref.~\cite{siddiki2004}. It is based on a local version of Ohm's
law, $\vec{j}(x)= \hat{\sigma}(x)\vec{E}(x)$ between current
density $\vec{j}(x)$ to  driving electric field $\vec{E}(x)$. The
position-dependent conductivity tensor $\hat{\sigma}(x)$ is taken
from a bulk calculation \cite{Ando82:437}, with  the electron
density $n_{\rm el}$ replaced by the local one, and a spatial
smoothening over a length of the order of the Fermi wave length is
performed, in order to avoid unphysical artifacts of a strictly
local transport model and of the TFA  \cite{siddiki2004}. Typical
results with a slightly modulated donor distribution
$n_0^B(x)=n_0^T(x)=n_0[1-0.01 \cos(5\pi x/2d)]$, introduced to
simulate the long-range-part of the impurity potentials
\cite{siddiki2004},
 are shown in
Fig.~\ref{fig:nonequdens}.

\section{Simulation of non-equilibrium\label{sec:simulation}}

The quasi-equilibrium model just introduced is, of course, not able to
describe hysteresis effects. It yields, however, qualitatively
correct results for the matched density case ($n^0_g=0$), where the
MRs for both layers are identical, and for the mismatched systems it
yields results in qualitative agreement with those found in the
``equilibrium measurements''. Moreover, the results shown in
Fig.~\ref{fig:nonequdens} give some hints towards the possible
origin of the hysteresis effects. We see that, in the plateau regime
of the QH effect, i.e., in the magnetic field interval in which
incompressible strips (ISs) exist \cite{siddiki2004},
 the position of the incompressible strips [local
filling factor $\nu(x)=2$] and the potential distribution change
drastically. To realize these changes, electrons must be
transported, e.g.\ by thermal activation, across the ISs. In real
samples (of finite length) this is extremely difficult, since
states at the Fermi level exist on both sides of an IS but not
within the IS, and since the region between ISs is not connected
to the contacts. Therefore, after a slight change of the magnetic
field, it may take an extremely long time until the electron
density relaxes to its new equilibrium distribution
\cite{jhuels2004,Zhu2000}.

\begin{figure}[t] {\centering
  \includegraphics[width=.6\linewidth,angle=-90]{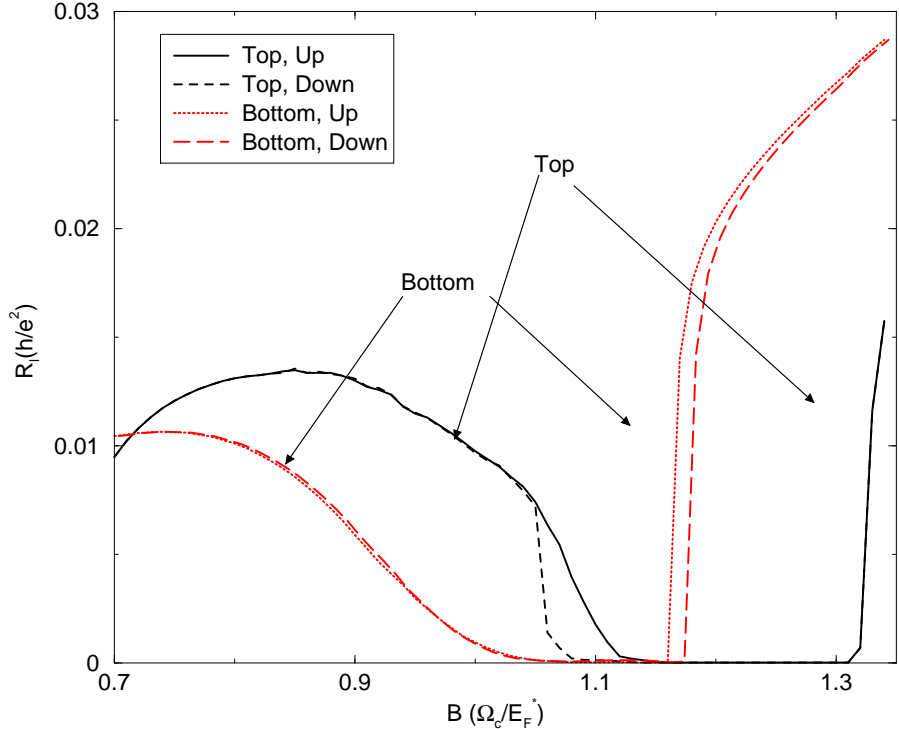}%
\caption{ \label{bare12} Longitudinal resistances for high density
mismatch parameter. The hysteresis-like effect is observed in the
active layer only if the passive layer is within the plateau
regime.
 }}
\end{figure}

To simulate this hindered approach to equilibrium,
we calculate the MRs of the current-carrying
 layer in a $B$-interval in which the other
layer exhibits a QH plateau, using different frozen-in
density and potential distributions for up- and down-sweeps
of the magnetic field $B$. For the down-sweep, we freeze in the
density profile of the other layer at the high-$B$ edge of
 its QH plateau and  for the
up-sweep we freeze it in at  the
low-$B$ edge, and use these fixed potentials to describe the other layer when
calculating the equilibrium and transport properties of the
current-carrying layer.
Thus, for the case shown in Fig.~\ref{fig:nonequdens}, we calculate
the MRs of the bottom layer in the QH plateau regime of the top layer
by fixing the potential due to the top layer at $\Omega_{c}/E^{*}_{F}
\approx 1.3$ for the down-sweep and at $\Omega_{c}/E^{*}_{F} \approx
1.0$ for the up-sweep. To calculate the MRs of the top layer in the
plateau regime of the bottom layer, we fix the potential profile
of the bottom layer at $\Omega_{c}/E^{*}_{F} \approx 1.16$ for the
down-sweep and at $\Omega_{c}/E^{*}_{F} \approx 1.05$ for the
up-sweep.
Since the MRs slightly outside the QH plateau regimes depend on the
electron and current density profiles \cite{siddiki2004}, we will
obtain slightly different results for the different sweep directions.
Figure~\ref{bare12} shows longitudinal magnetoresistances calculated
along such lines.

\section{Summary}
We report on measurements and calculations of the magnetoresistances
in separately contacted {\em e-e} bilayer
systems. The calculation, based on a self-consistent
Thomas-Fermi-Poisson theory of the equilibrium state and a quasi-local
transport theory \cite{siddiki2004}, yields reasonable agreement with
the measurements for density-matched layers and for mismatched
layers, if the measurement allows for thermal relaxation
 at each value of the magnetic field $B$. $B$-sweeps at constantly low
temperature show hysteresis between up and down sweep in the
current-carrying layer, provided the other layer exhibits the
quantized Hall effect. A model, based on the existence of
incompressible regions in the quantum Hall states \cite{siddiki2004}
 and the extremely
long relaxation times in compressible regions surrounded by
incompressible ones \cite{jhuels2004}, is worked out and can explain
the observed hysteresis effects.


\begin{thebibliography}{10}

\bibitem{Ahlswede01:562}
E. Ahlswede {\it et~al.}, Physica B {\bf 298},  562  (2001).

\bibitem{Ahlswede02:165}
E. Ahlswede, J. Weis, K. von Klitzing, and K. Eberl, Physica E
{\bf 12},  165
  (2002).

\bibitem{Guven03:115327}
K. G{\"u}ven and R.~R. Gerhardts, Phys. Rev. B {\bf 67},  115327
(2003).

\bibitem{siddiki2004}
A. Siddiki and R.~R. Gerhardts, Phys. Rev. B {\bf 70},  195335
(2004).

\bibitem{Chklovskii92:4026}
D.~B. Chklovskii, B.~I. Shklovskii, and L.~I. Glazman, Phys. Rev.
B {\bf 46},
  4026  (1992).

\bibitem{Siddiki03:125315}
A. Siddiki and R.~R. Gerhardts, Phys. Rev. B {\bf 68},  125315
(2003).

\bibitem{Gramila91:1216}
T.~J. Gramila {\it et~al.}, Phys. Rev. Lett. {\bf 66},  1216
(1991).

\bibitem{Zheng93:8203}
L. Zheng and A.~H. MacDonalds, Phys. Rev. B {\bf 48},  8203
(1993).

\bibitem{Bonsager97:10314}
M.~C. Bonsager, K. Flensberg, B.~Y.~K. Hu, and A.~P. Jauho, Phys.
Rev. B {\bf
  56},  10314  (1997).

\bibitem{tutuc}
E. Tutuc {\it et~al.}, Phys. Rev. B {\bf 68},  201308  (2003).

\bibitem{pan}
W. Pan, J. Reno, and J. Simmons, cond-mat/0407577  (2004).

\bibitem{Eisenstein90:2324}
J.~P. Eisenstein, L.~N. Pfeiffer, and K.~W. West, Appl. Phys.
Lett. {\bf 57},
  2324  (1990).

\bibitem{Ando82:437}
T. Ando, A.~B. Fowler, and F. Stern, Rev. Mod. Phys. {\bf 54},
437  (1982).

\bibitem{jhuels2004}
J. Huels {\it et~al.}, Phys. Rev. B {\bf 69},  085319  (2004).

\bibitem{Zhu2000}
J. Zhu {\it et~al.}, Phys. Rev. B {\bf 60},  5536  (1999).

\end{thebibliography}

\end{document}